\documentclass[a4paper,12pt]{article}
\pdfoutput=1 % if your are submitting a pdflatex (i.e. if you have
\usepackage{jinstpub} % for details on the use of the package, please
\usepackage[active,new,old]{correct}
\usepackage{graphicx,epsfig}

\title{\boldmath Electron channeling experiments with bent silicon single crystals - a reanalysis based on a modified Fokker-Planck equation}

\author[]{H. Backe}
\affiliation[]{Institute for Nuclear Physics of Johannes Gutenberg-University, Johann-Joachim-Becher-Weg 45 \\D-55128 Mainz, Germany}
\emailAdd{backe@uni-mainz.de}

\abstract{A surprising small dechanneling length was observed at (111) channeling of ultrarelativistic electrons in a 60 $\mu$m thick silicon single crystal with a bending radius of 0.15 m. The experiments were conducted at beam energies between 3.35 and 14 GeV at the Facility for Advanced Accelerator Experimental Tests (FACET at SLAC, USA). It is shown in this paper that the small dechanneling lengths can well be reproduced with a modified Fokker-Planck equation for plane crystals in
which a crystal bending has been heuristically introduced. Encouraged by this
result experiments have been reconsidered which were performed at the Mainz Microtron MAMI with (110) silicon undulator crystals. The results obtained with the modified Fokker-Planck equation suggest that the observed rather low undulator peak intensity originates from the strongly reduced dechanneling length of electrons in the bent sections of the undulator. A scaling law derived on the basis of the modified Fokker-Planck equation reveals
optimized parameters of electron based undulators as possible radiation sources in the $X$- and $\gamma$-ray region.}


\proceeding{XII$^{\text{th}}$ International Symposium, RREPS-17\\
  September 18-22, 2017\\
  Hamburg, Germany}

\begin{document}
\maketitle
\flushbottom

\section{Introduction} \label{sec:intro}
The phenomenon of channeling of positively and negatively charged particles plays an important role in the high energy physics domain like for beam bending and collimation \cite{Car08}. Another intriguing field of research is the development of crystalline undulators for intense radiation production with energies of 100 keV or higher employing relativistic positrons or electrons with energies in the order of a few hundred MeV or more. Such devices were proposed a long time ago \cite{KapK80, BarD80} and
theoretically investigated in detail more recently \cite{KorS04,
BelM06, Bar12, KorS13}. A very important prerequisite for an
experimental realization of such undulators is the knowledge of
the dechanneling length, i.e., the length in which the charged
particle remains in an periodically bent crystal in the channel.
Of particular interest are electrons since high quality electron
beams can much easier be produced as compared to positrons. Currently simulation calculations are widely used to get information on dechanneling lengths, see, e.g., the recent article of A. V. Korol, A. V. Solov'yov et al. \cite{KorB16}. However, also such types of calculations need experimental verifications.

There are principally two possibilities to measure the dechanneling length.
The first one is based on a variation of the crystal thickness. However, the results
are of little relevance for the goal to construct a crystal undulator
radiation source for which the dechanneling length in bent crystals is of
interest. Fortunately, there exists a very elegant second possibility, namely to observe dechanneled electrons from a bent crystal. Experiments have been performed at the Facility for Advanced Accelerator
Experimental Tests (FACET) at SLAC using a bent silicon single crystal for channeling
in the (111) planes at beam energies between 3.35 and 14 GeV \cite{MazB14, WisU16}. In this paper some of these data will be explained employing the Fokker-Planck equation which has been heuristically modified for bending of the crystal. The aim was to verify a procedure in order to predict dechanneling lengths for the (110) plane of silicon and diamond undulator crystals for experiments at the Mainz microtron facility MAMI at beam energies below 855 MeV.

In the next subsection first the modified Fokker-Planck equation will be
described. This section is followed by a comparison of the experimental dechanneling length results of T. N. Wistisen, U. I. Uggerh{\o}j, U. Wienands et al. \cite{WisU16} obtained at FACET (SLAC) with calculations on the basis of the modified Fokker-Planck equation. Encouraged by the good agreement between calculations and experiment a section follows in which dechanneling in bent (110) planes of silicon single crystals will be investigated and compared with previous undulator experiments at MAMI \cite{BacK12, BacK12A}. Finally the question will be addressed which kind of radiation features can be expected from an optimized large amplitude undulator operating with electrons.

%%%%%%%%%%%%%%%%%%%%%%%%%%%%%%%%%%%%%%%%%%%%%%%%%%%%%%%%%%%%%%%%
\begin{figure}[tb]
%\hspace{3.0 cm}
\centering
    \includegraphics[angle=0,scale=0.5,clip] {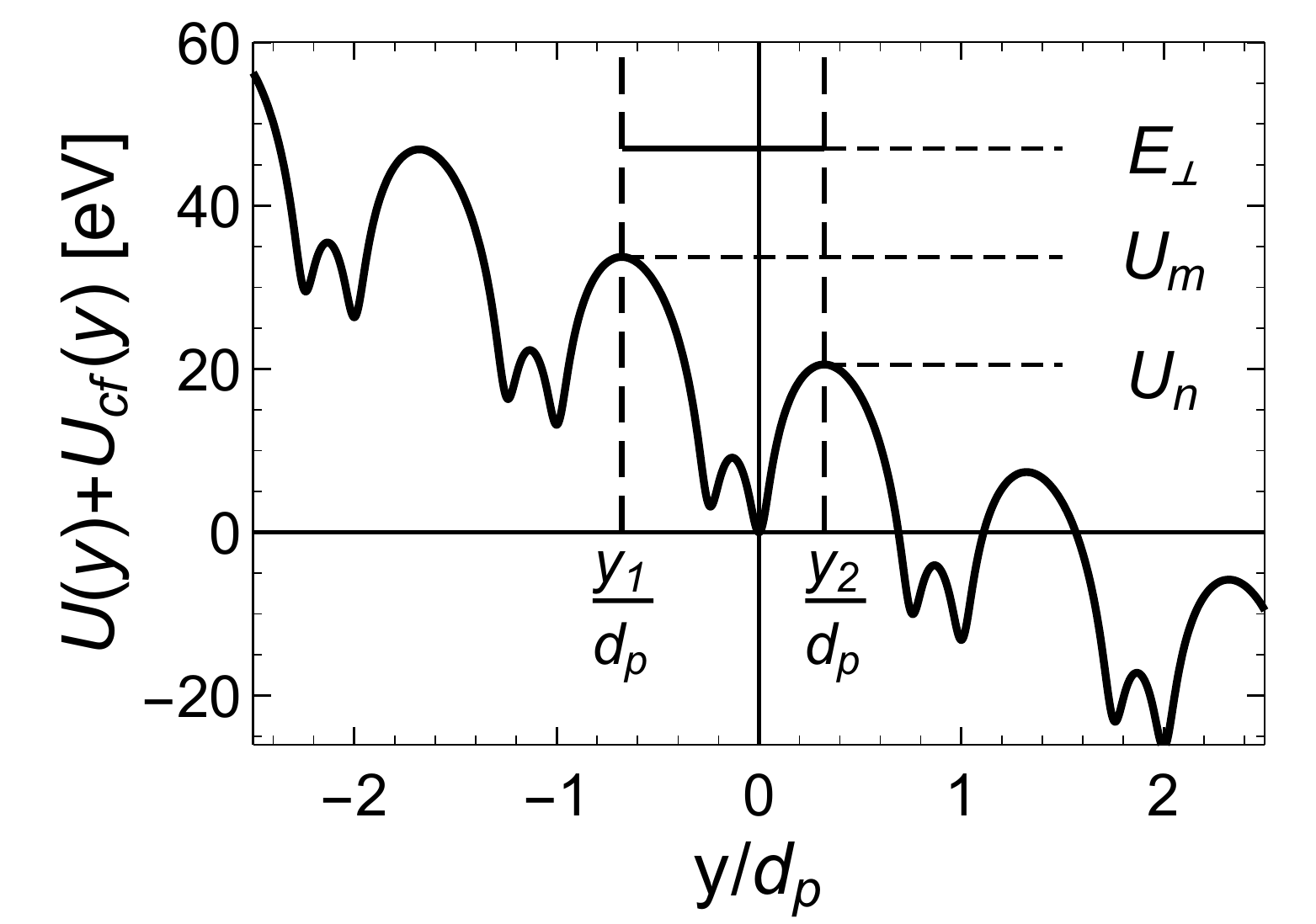}
\caption[] {Effective potential $U_{eff} = U(y)+U_{cf}(y)$ with $U(y)$ the potential of the (111) planes of a silicon
single crystal and $U_{cf}(y)$ the centrifugal potential
superimposed. Bending radius $R$ = 0.15 m, electron beam energy
6.3 GeV, interplanar distance $d_p$ = 3.135 \AA~, maximum $U_m$ =
33.71 eV at $y_1/d_p$, maximum $U_n$ = 20.54 eV at $y_2/d_p$. The
transverse energy $E_{\perp}$ is schematically also
indicated.} \label{Potential}
\end{figure}
%%%%%%%%%%%%%%%%%%%%%%%%%%%%%%%%%%%%%%%%%%%%%%%%%%%%%%%%%%%%%%%%

\section{The modified Fokker-Planck equation for a bent crystal}
\label{modifiedFokkerPlanck} In figure
\ref{Potential} an example of the effective potential $U_{eff}(y)=U(y)+U_{cf}(y)$  is shown in which
a modified Fokker-Planck equation must be solved. The figure depicts a superposition of the
potential $U(y)$ for an electron channeling in a straight silicon single
crystal with the centrifugal potential
\begin{eqnarray}\label{centrifugalPotential}
U_{cf}(y)=-\frac{\gamma m_e c^2 \beta^2}{R} y = - \frac{p v}{R} y
\end{eqnarray}
which accounts for the bending. Here are $\gamma =
1/\sqrt{1-\beta^2}$ the Lorentz factor, $\beta = v/c$, $v$ the
electron velocity, $c$ the speed of light, $m_e$ the rest mass of
the electron, $p$ its momentum, and $R$ the bending radius. A
coordinate system has been chosen with the $x$ axis pointing into the
initial beam direction, and the $y$ axis perpendicular to the channeling plane. The modified Fokker-Planck equation, with $F(x, E_{\perp}) = \Delta P/\Delta E_{\perp}$ the probability $\Delta P$ per transverse energy interval $\Delta E_{\perp}$, and  $J(x, E_{\perp}) = J_{diff}(x, E_{\perp})+J_{drift}(x, E_{\perp})$ the probability current, reads
\begin{eqnarray}
\frac{\partial F(x, E_{\perp})}{\partial x}+\frac{\partial J(x, E_{\perp})}{\partial E_{\perp}} + \theta ({E_ \bot } - {U_n}) \frac{\partial J_{drift}^{(cf)}(x, E_{\perp})}{\partial E_{\perp}}=0~, \label{FokkerPlanck1}\\
J(x, E_{\perp}) = -\frac{\partial}{\partial E_{\perp}} \bigg[D_{diff}^{(2)}(E_{\perp})F(x,E_{\perp})\bigg]+D_{drift}^{(1)}(E_{\perp}) F(x, E_{\perp})~, \label{FokkerPlanck2}\\
J_{drift}^{(cf)}(x, E_{\perp}) = \frac{1}{R}\sqrt{2(E_{\perp}-U_{n})pv}\cdot F(x, E_{\perp})~. \label{FokkerPlanck3}
\end{eqnarray}
An additional drift current term $J_{drift}^{(cf)}(x, E_{\perp})$ has
been heuristically introduced which accounts for the motion of the
probability density $F(x,E_{\perp})$ due to the centrifugal force. It acts
directly only on continuum states with $E_{\perp} > U_{n}$. This fact has
been expressed by the Heaviside $\theta$ function in equation
(\ref{FokkerPlanck1}). How this term comes about will be explained below.

Details of the basic underlying formalism for the Fokker-Planck equation at
planar channeling in straight crystals are described in, e.g.,
\cite{BaiK98, BelT81, KumK89} and references cited therein and
will not be repeated here. It should only be mentioned that the Fokker-Planck
equation in phase space has been simplified assuming statistical
equilibrium meaning that the probability distribution of the
particle in the channel is assumed to be representable by
\begin{eqnarray}\label{ProbabilityDistribution}
\frac{dP}{dy}(y, E_{\perp})=\frac{2}{T(E_{\perp})}\sqrt{\frac{\gamma
m_e}{{2\big(E_{\perp}}-U_{eff}(y)\big)}}
\end{eqnarray}
with a time period
\begin{eqnarray}\label{Period}
T(E_{\perp}) = 2\int\limits_{y_{min}}^{y_{max}} \sqrt{\frac{\gamma
m_e}{{2\big(E_{\perp}}-U_{eff}(y)\big)}}~~dy.
\end{eqnarray}
The time parameter $T(E_{\perp})$ is the period for one full cycle
for a bound state and twice the transit time over the channel for
a free state. The limits of the integral $y_{min}$ and $y_{max}$
are roots of the equation $U_{eff}(y) = U(y)+U_{cf}(y) = E_{\perp}$ with $y_{max} = y_2$
for $E_{\perp}>U_n$, and $y_{min} = y_1$ for $E_{\perp}>U_m$.

The channeling potential $U(y)$ for the straight crystal has been
calculated in the Moli\`{e}re approximation, see e.g. Baier et al. \cite[Ch.
9.1 and equation (7.45)]{BaiK98}. The electron distribution in the
channeling planes has been neglected. Without loss of generality
it can be assumed that only the potential in one period, indicated by the
vertical dashed lines in figure \ref{Potential}, must be considered
for the calculations of the drift coefficient
\begin{eqnarray}\label{DriftCoefficient}
D_{drift}^{(1)}(E_{\perp})= \bigg\langle \overline{\frac{\Delta
E_{\perp}}{\Delta x}}\bigg\rangle _{T}~,
\end{eqnarray}
and the diffusion coefficient
\begin{equation}\label{DiffusionCoefficient}
D_{diff}^{(2)}(E_{\perp}) =
\frac{1}{2}\bigg\langle\frac{\overline{(\Delta
E_{\perp})^2}}{\Delta x} \bigg\rangle_{T} = \bigg\langle
2\overline{\frac{\Delta E_{\perp}}{\Delta x}}~\Big
(E_{\perp}-U_{eff}(y)\Big) \bigg\rangle _{T}~.
\end{equation}
Both quantities are mean values with respect to the distribution
function equation (\ref{ProbabilityDistribution}) over one period.

The drift coefficient $D_{drift}^{(1)}(E_{\perp})$ has been calculated
in the Kitagava-Ohtsuki approximation \cite{KitO73}, which
accounts for phonon excitation, by means of the integral
\begin{eqnarray}\label{KitagavaOhtsuki}
D_{drift}^{(1)}(E_{\perp}) = \frac{E_{s}^{2}}{2pv X_{0}}
\frac{2}{T(E_{\perp})c}
\frac{1}{2}\sum_{i=1}^{2}\int\limits_{y_{min}}^{y_{max}}\frac{d_{p}}{\sqrt{2\pi}u_{1}}
\frac{\exp(-(\eta-y_i^{plane})^{2}/2u_{1}^{2})}
{\sqrt{2\big(E_{\perp}-U_{eff}(\eta)\big)/\gamma m_{e} c^{2}}}~
d\eta.
\end{eqnarray}
A modified parameter $E_s$ = 10.6 MeV has been used in the standard
deviation of the Gaussian scattering distribution in the Rossi-Greisen approximation \cite[\S 22 "Multiple Scattering...."]{RosG41} which reads
\begin{equation}\label{ScatteringAngle}
\theta_{0}=\frac{10.6 ~\mbox{MeV}}{pv} \sqrt{\frac{x}{X_{0}}}~,
\end{equation}
for the reasoning see \cite[footnote in chapter 3]{BacL15}. The
quantity $X_{0}$ = 0.0936 m is the radiation length of silicon, and $u_{1}$ = 0.076 \AA~ the standard deviation of
the thermal vibration amplitude. The quantity $d_p = a/\sqrt{3}$ =
3.135 \AA~ in the nominator of the integral is the interplanar
distance, with $a$ = 5.431 \AA~ the lattice constant of silicon at
300 K. It should be mentioned that in case of (111) geometry two planes account for the potential minima which
are $\Delta^{pot}$ = 0.7515 \AA~ apart. The
lattice planes have a distance of $\Delta^{plane} = a/(4
\sqrt{3})$ = 0.7839 \AA. Since the minimum of the potential
$U(y)$ was shifted to $y$ = 0, the $y_i^{plane}$ read
$y_1^{plane} = (\Delta^{plane}-\Delta^{pot})/2$, and $y_2^{plane} =
-(\Delta^{plane}+\Delta^{pot})/2$. The factor 1/2 in front of the
sum in equation (\ref{KitagavaOhtsuki}) accounts for a proper
normalization since each of the exponential functions is normalized to
one.

\begin{figure}[tbh]
%\vspace*{2cm}
\centering
    \begin{minipage}[t]{0.45\linewidth}
    %\centering
    \hspace*{-0.5cm}
    \includegraphics[angle=0,scale=0.57,clip]{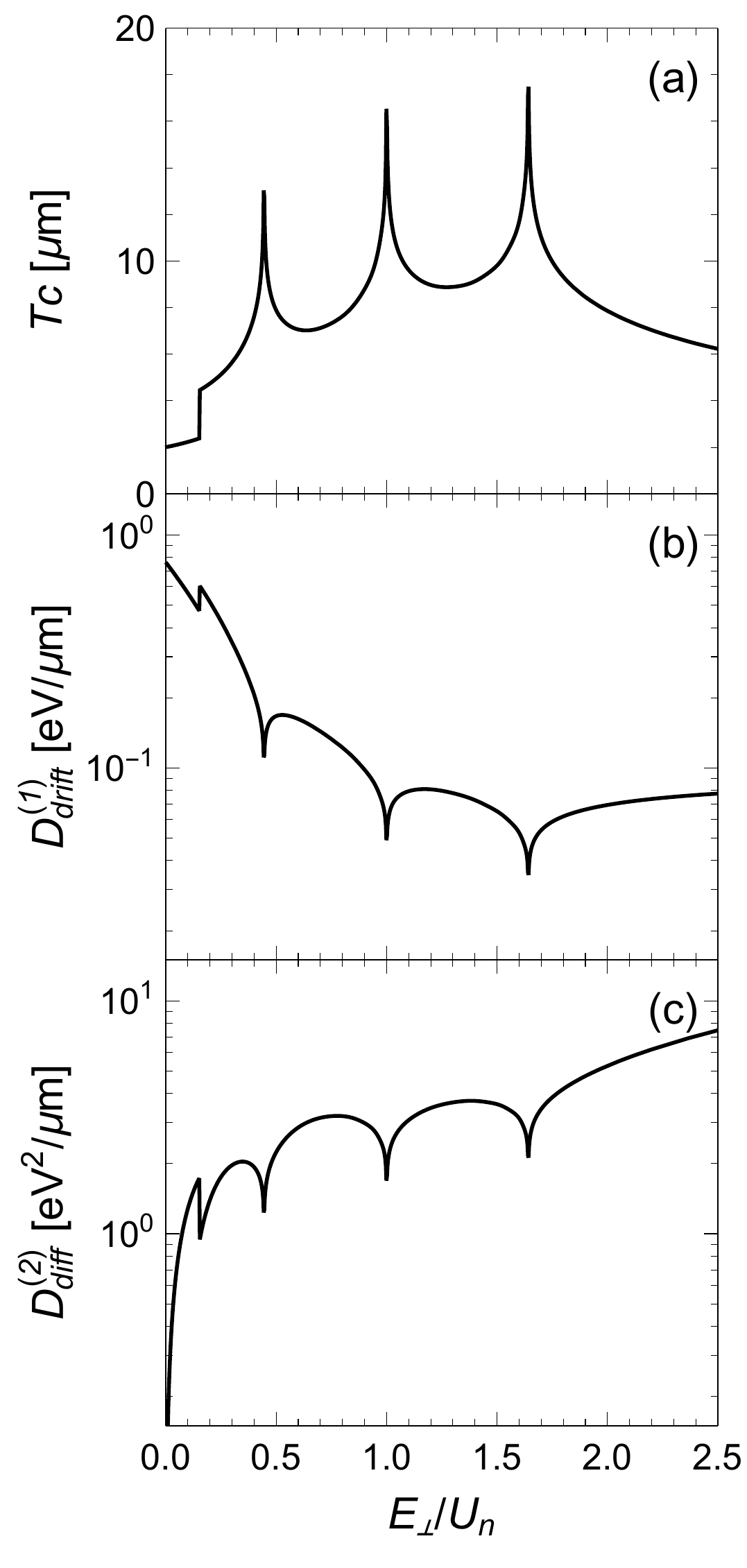}
    \end{minipage}
    %\hfill
    \begin{minipage}[b]{0.45\linewidth}
    \centering
    \includegraphics[angle=0,scale=0.57,clip]{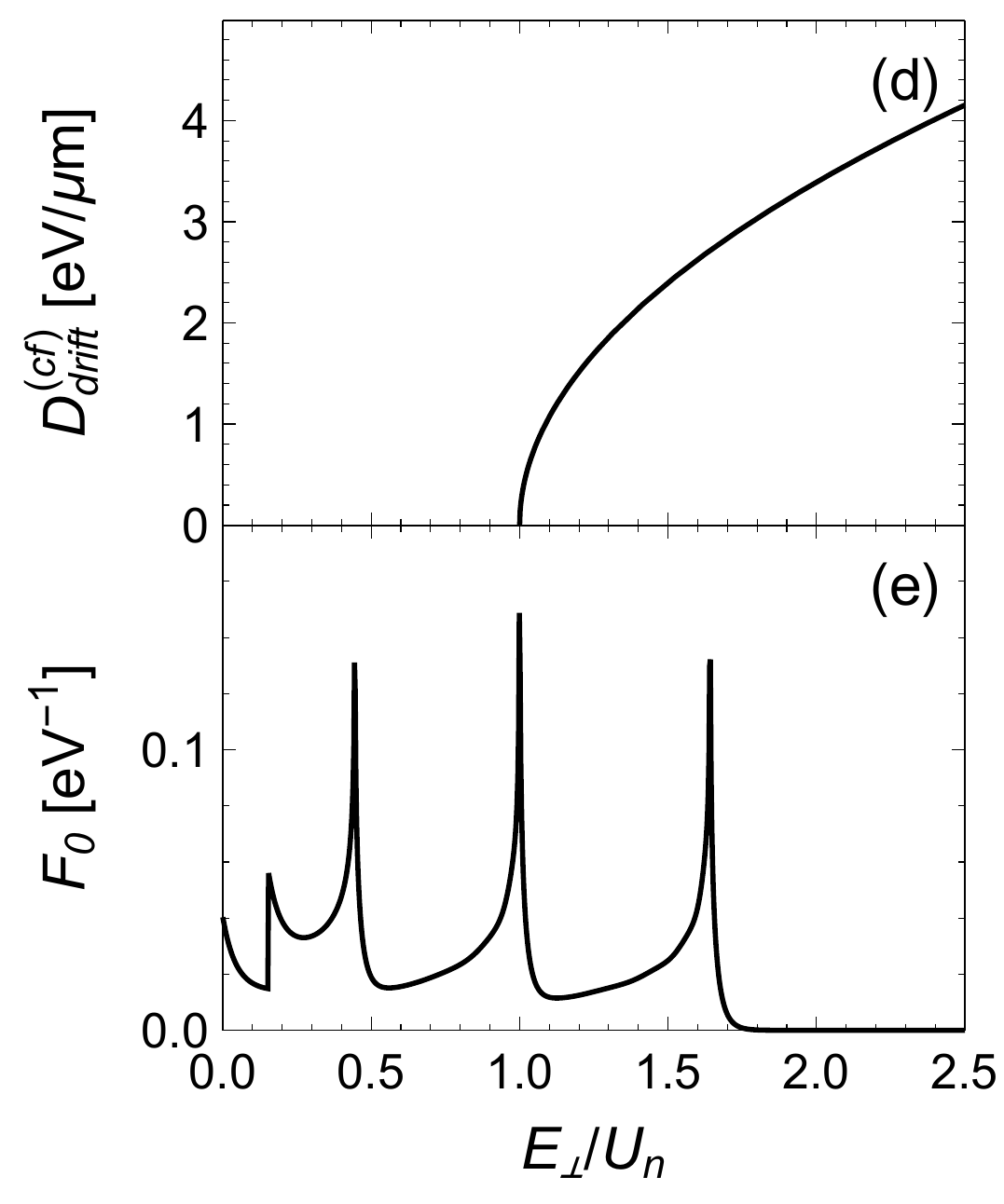}
    \includegraphics[angle=0,scale=0.53,clip]{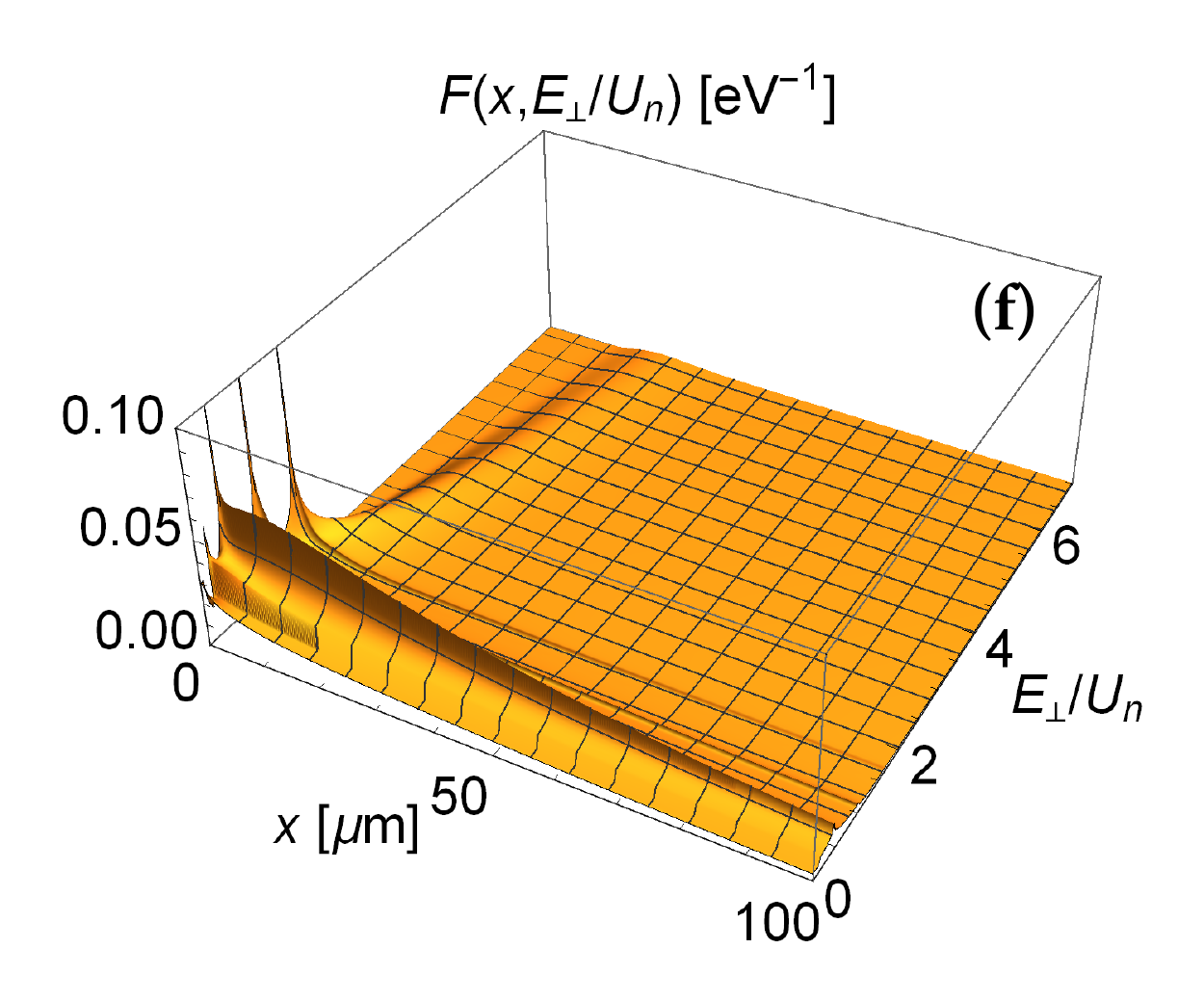}
    \end{minipage}
%\vspace*{-2.1cm}
\caption[]{(a) Time parameter $Tc$,  (b) drift coefficient
$D_{drift}^{(1)}$, (c) diffusion coefficient $D_{diff}^{(2)}$, (d) drift
coefficient $D_{drift}^{(cf)}$ due to centrifugal force, (e)
initial distribution $F_0$ at $x = 0$ all as function of
normalized transverse energy $E_{\perp}/U_n$, and (f) numerical solution $F(x, E_{\perp}/U_n)$ of the modified Fokker-Planck equation. Calculations for beam energy $E$ = 6.3 GeV, bending radius $R$ = 0.15 m, standard deviation $\sigma '_{y}=10~\mu$rad for angular divergence of the electron beam, and $U_n$ = 20.54 eV.} \label{TdEdxEtc}
\end{figure}

The diffusion coefficient $D_{diff}^{(2)}(E_{\perp})$ has been
calculated by means of the equation
\begin{equation}\label{DriftDiffusion}
\frac{\partial}{\partial E_{\perp}} \Big[T(E_{\perp})\cdot
D_{diff}^{(2)}({E_{\perp})}\Big]= T(E_{\perp})\cdot
D_{drift}^{(1)}(E_{\perp}),
\end{equation}
i.e., by integration of the known right hand side of this equation and division by
$T(E_{\perp})$. Examples for $T(E_{\perp})$,
$D_{drift}^{(1)}(E_{\perp})$, and $D_{diff}^{(2)}(E_{\perp})$ are shown
in figure \ref{TdEdxEtc} (a-c).
%%%%%%%%%%%%%%%%%%%%%%%%%%%%%%%%%%%%%%%%%%%%%%%%%%%%%%%%%%%%%%%%
%%%%%%%%%%%%%%%%%%%%%%%%%%%%%%%%%%%%%%%%%%%%%%%%%%%%%%%%%%%%%%%%

Let us turn to the question how the additional drift current term in equation
(\ref{FokkerPlanck1}) comes about. In principle, the probability density
$F(x, y, E_{\perp})$ depends on three variables. However, a particle which
enters into the continuum at $x = x_0$ moves, viewed from the bent channel which is assumed to be straight, on
a circle with radius $R$ which can be approximated by the equation
$y=(x-x_0)^2/2R$. The transverse energy $E_{\perp} = U_n + pv/R\cdot y$
is translational invariant with respect to $x_0$, i.e., $E_{\perp}$ evolves as function of $y$ irrespective of the depth $x_0$ at which the particle leaves the channel. Therefore the variable $y$ can be
eliminated leading to $E_{\perp} = U_n +(pv/R)\cdot (x-x_0)^2/2R$. The
additional probability current is expressed as $J_{drift}^{(cf)}(x,
E_{\perp}) = \partial E_{\perp}(x)/\partial x \cdot F(x, E_{\perp})$ which
leads with $\partial E_{\perp}(x)/\partial x = pv (x-x_0)/R^2$, and $x-x_0 =
R \sqrt{2(E_{\perp}-U_n)/pv}$ to equation (\ref{FokkerPlanck3}).

One may introduce the drift coefficient
\begin{eqnarray}\label{DriftCoeffCF}
D_{drift}^{(cf)}(E_{\perp})=\frac{1}{R}\sqrt{2(E_{\perp}-U_{n})\cdot pv}
\end{eqnarray}
which accounts for the motion of the probability density $F(x,
E_{\perp})$ due to the centrifugal force. It is shown in figure \ref{TdEdxEtc} (d). The additional drift coefficient does not act on bound states with $E_{\perp} < U_{n}$.

For the initial conditions at $x=0$, required for the numerical
solution of the Fokker-Planck equation, a uniform distribution of
the electron across the transverse $y$ coordinate, and a Gaussian
scattering distribution tilted by an angle $\psi^0_{in}$, and with
standard deviation $\sigma '_{y}$ for the angular divergence were
assumed. For a function with the two random variables $y_{in}$ and
$\psi_{in}$, which are connected to $E_{\perp}$ by the relation
$E_{\perp}/U_{0} =
U_{eff}(y_{in})/U_{0}+(\psi_{in}/\psi_{c}-\psi^0_{in}/\psi_{c})^2$, with
$\psi_{c} = \sqrt{2 U_0/\gamma m_e c^2}$ the critical angle for a
potential depth $U_0$ = 25.15 eV, the formalism described in Ref.
\cite[chapter 6]{Pap89} was applied. This approach leads to the
probability density
\begin{eqnarray}\label{initialProbability}
F_0(E_{\perp})~  =  \frac{1}{2\sqrt{2\pi} (\sigma
'_{y}/\psi_{c}) U_{0}}\int\limits_{y_{min}}^{y_{max}} \Bigg(\frac{\exp\bigg[
-\frac{\big(\sqrt{(E_{\perp}-U_{eff}(\eta))/U_{0}}+
\psi_{in}^0/\psi_c\big)^2}{2(\sigma '_{y}/\psi_{c})^{2}} \bigg]}{
\sqrt{(E_{\perp}-U_{eff}(\eta))/U_{0}}}~+
\nonumber\\
& &\hspace{-7 cm} + \frac{\exp\bigg[
-\frac{\big(-\sqrt{(E_{\perp}-U_{eff}(\eta))/U_{0}}+
\psi_{in}^0/\psi_c\big)^2}{2(\sigma '_{y}/\psi_{c})^{2}} \bigg]}{
\sqrt{(E_{\perp}-U_{eff}(\eta))/U_{0}}}\Bigg)~d\eta.
\end{eqnarray}
The initial distribution with parameters described in the caption is shown in figure
\ref{TdEdxEtc} (e). A fraction of 64.5 \% is captured in a bound
state with $E_{\perp} < U_n$.

The probability density $F(x, E_{\perp})$ of a free particle moves as function of $x$ rapidly
into the $E_{\perp}$ direction while preserving normalization. This fact can
be seen in figure \ref{TdEdxEtc} (f) by the ridge which originates from the
primary unbound density at $E_{\perp}/U_n \approx 1.64$ of the initial distribution, see figure \ref{TdEdxEtc} (e). While the particle moves over the region with $y>y_2$, see figure \ref{Potential}, volume capture may happen which is not included in the calculations. For $E_{\perp}~>~U_m$ the particle may also enter via the diffusion term in the Fokker-Planck equation into the region with $y<y_1$ in which it may also experience volume capture (rechanneling). This possibility has been neglected as well. Therefore, the results for the dechanneling length calculations represent lower limits which should be taken in mind when compared with experimental results.

\section{Comparison of dechanneling length measurements with results of the modified Fokker-Planck equation for bent crystals} \label{dechannelingAnalysis}
\begin{figure}[tbh]
%\vspace*{-1.1cm}
\centering
    \begin{minipage}[t]{0.45\linewidth}
     \hspace{-0.5cm}
    %\centering
    \includegraphics[angle=0,scale=0.56,clip]{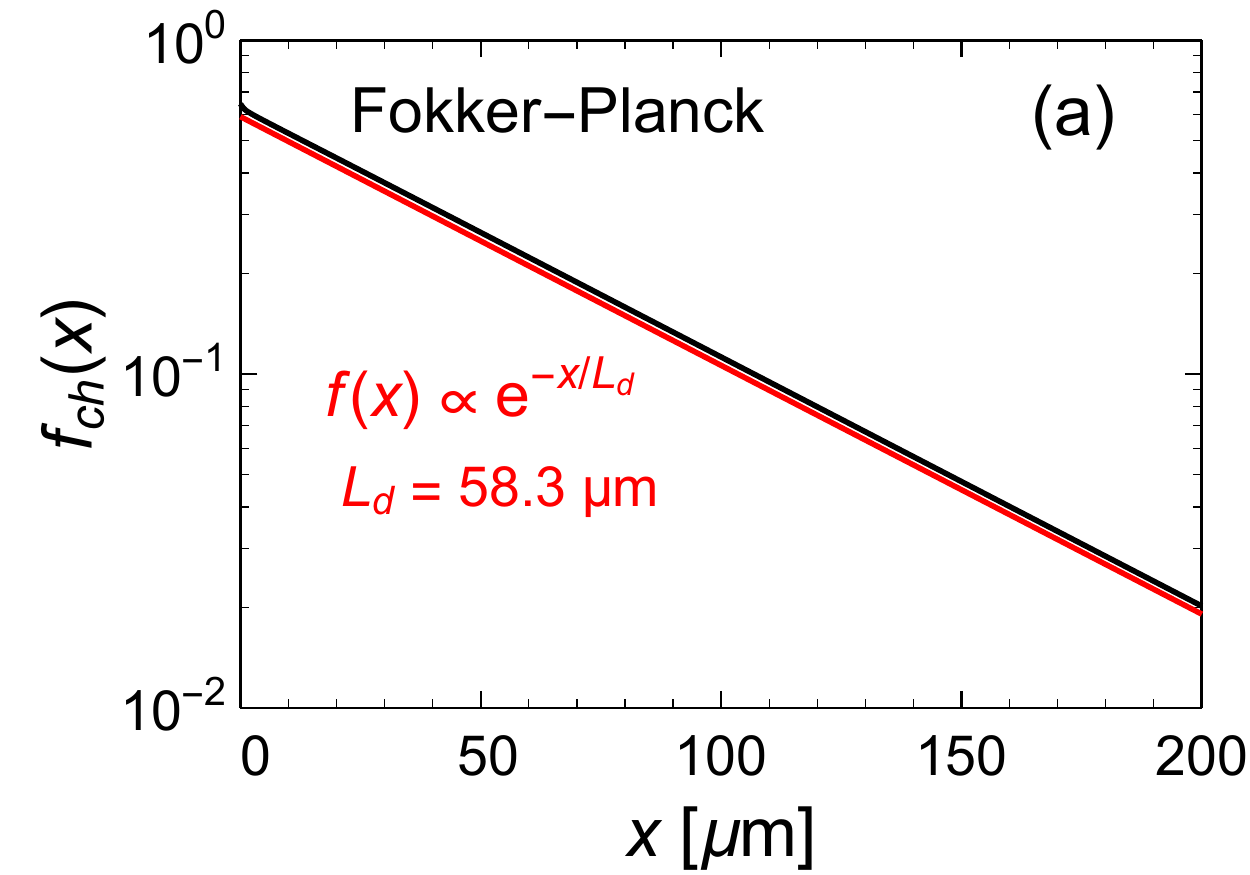}
    \end{minipage}
    %\hfill
    \hspace{0.3cm}
    \begin{minipage}[t]{0.45\linewidth}
    %\centering
    \includegraphics[angle=0,scale=0.56,clip]{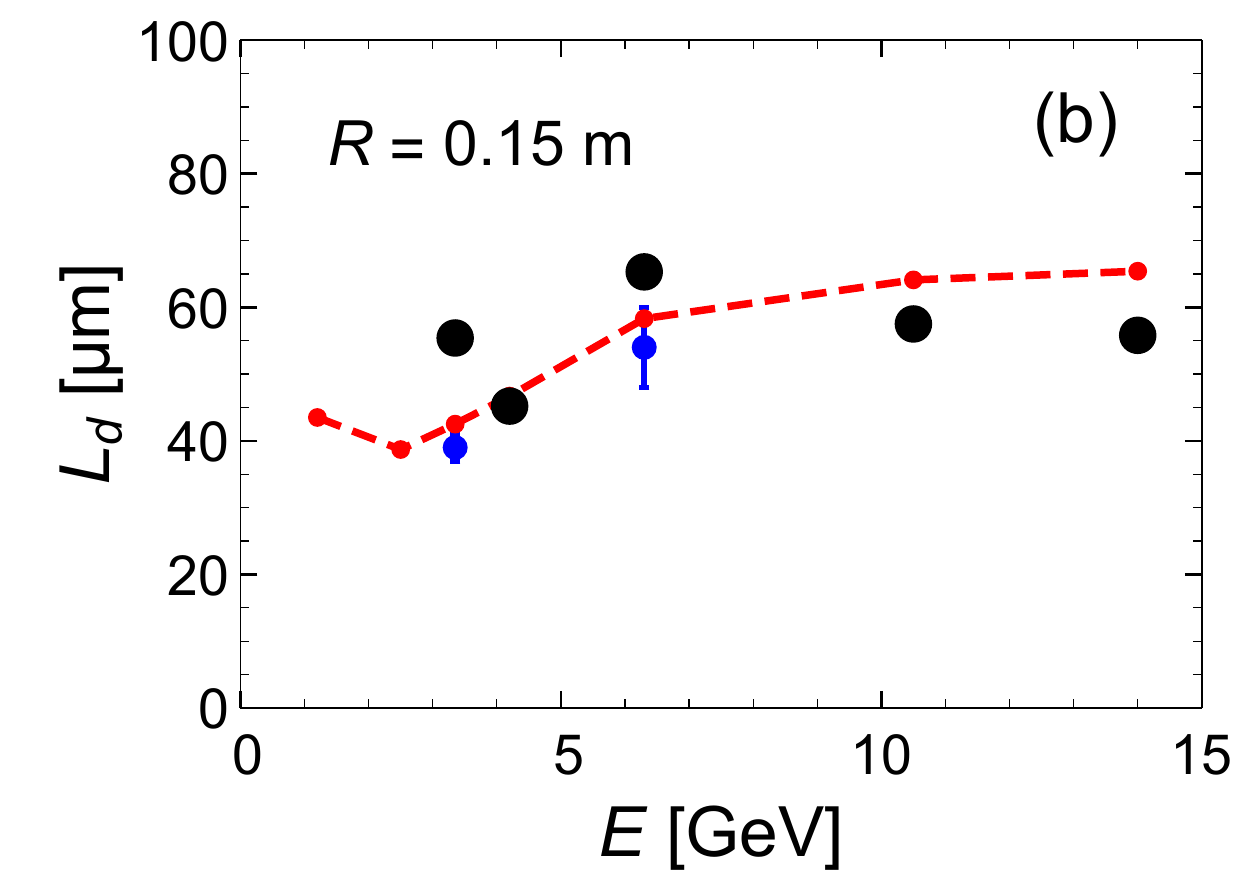}
    \end{minipage}
%\vspace*{-2.1cm}
\caption[]{(a) Solution of the modified Fokker-Planck equation (black line) compared with an exponential decay (red line). Shown is the fraction of electrons captured in the potential pocket $f_{ch}(x)$ as function of the distance $x$ traversed by the electron. The dechanneling length $L_{d}$ was adapted. Beam energy 6.3 GeV. (b) Comparison of the  dechanneling lengths obtained this way (red points) with experimental results (full circles) \cite{WisU16}. The red dashed lines connect the calculated red points to guide the eye. The blue error bars are simulation calculations of G. B. Sushko , A. V. Korol, A. V. Solov$'$yov \cite{SusK15} on the basis of the MBN Explorer software package.} \label{LdechExpFPSi111}
\end{figure}

As already mentioned in the Introduction, dechanneling length measurements have been performed at FACET (SLAC) for Si (111) planes at beam energies between 3.35 and 14 GeV \cite{WisU16}. The crystal thickness and the bending radius $R$ were 60 $\mu$m and 0.15 m, respectively.
In figure \ref{LdechExpFPSi111} (a) an example of calculations on the basis of the modified Fokker-Planck equation (\ref{FokkerPlanck1}-\ref{FokkerPlanck3}) is shown. It can be seen that the occupation of the potential pocket $f_{ch}(x)$ can well be approximated by an exponential function. With the exception of an unimportant scaling factor no further parameters were adapted. The dechanneling lengths $L_{d}$ obtained this way are compared in figure \ref{LdechExpFPSi111} (b) with the experimental results \cite{WisU16}. The gross features of the measurements are well described. Shown are also calculations on the basis of the MBN Explorer software package \cite{SusK15}.

%%%%%%%%%%%%%%%%%%%%%%%%%%%%%%%%%%%%%%%%%%%%%%%%%%%%%%%%%%%%%%%%
%%%%%%%%%%%%%%%%%%%%%%%%%%%%%%%%%%%%%%%%%%%%%%%%%%%%%%%%%%%%%%%%
%%%%%%%%%%%%%%%%%%%%%%%%%%%%%%%%%%%%%%%%%%%%%%%%%%%%%%%%%%%%%%%%
\begin{figure}[tbh]
%\vspace*{-1.1cm}
\centering
   \includegraphics[angle=0,scale=0.6,clip]{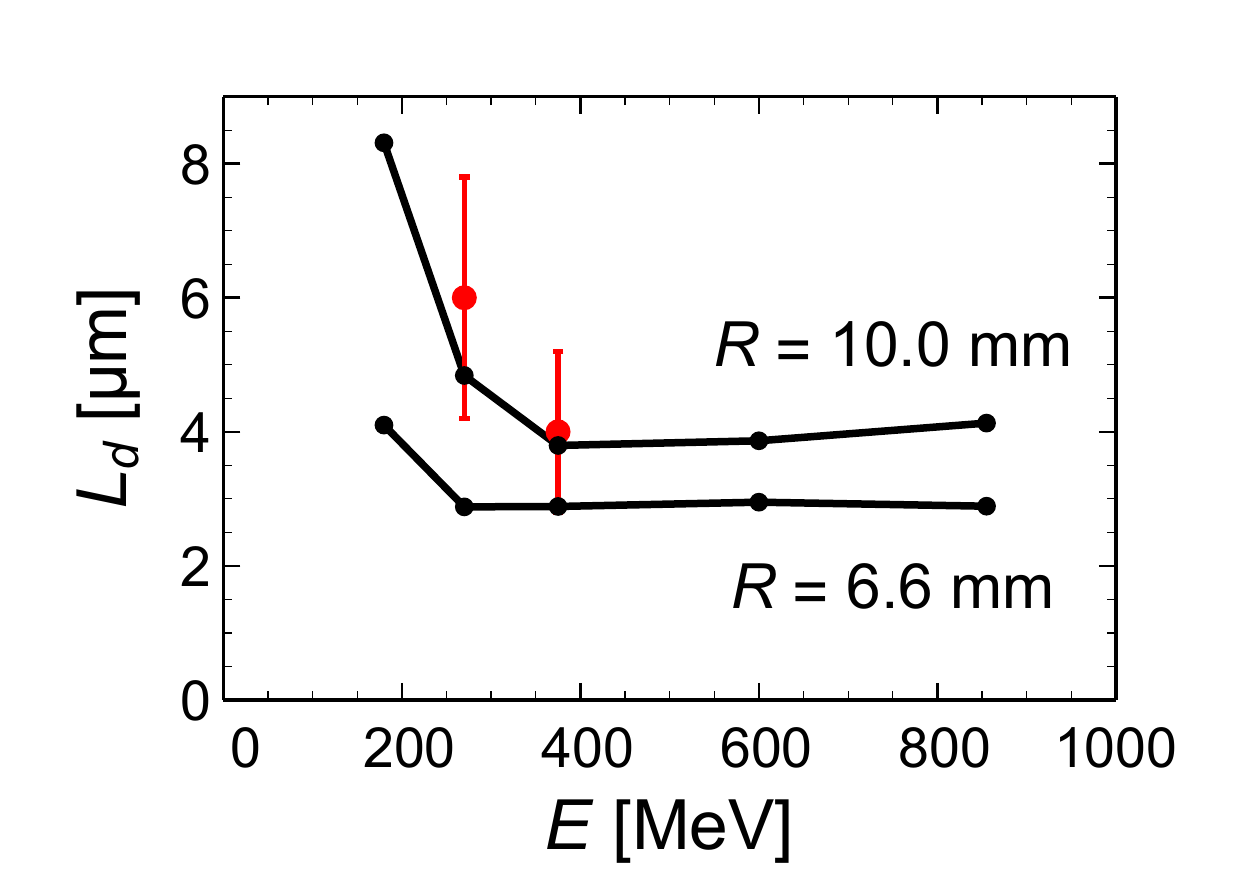}
%\vspace*{-2.1cm}
\caption[]{Comparison of the calculated dechanneling length as
function of the beam energy with results extracted from experiments at MAMI \cite{BacK12, BacK12A}. The design value of the bending radius was $R$ = 6.6 mm. Shown are also calculations for $R$ = 10 mm which seem to be in better agreement with experiments (red error bars).} \label{Si110undulatorR6p6MAMI}
\end{figure}
%%%%%%%%%%%%%%%%%%%%%%%%%%%%%%%%%%%%%%%%%%%%%%%%%%%%%%%%%%%%%%%%

Encouraged by this result calculations have been performed for a Si (110) four period undulator with which experiments were performed at MAMI beam energies of 270 and 375 MeV \cite{BacK12, BacK12A}. The design values were $\lambda_U = 9.9~\mu$m for the period length, and $A = 4.64$ \AA~ for the amplitude, resulting in a bending radius $R \cong (\lambda_U/4)^2/2A$ = 6.6 mm. In figure \ref{Si110undulatorR6p6MAMI} the solutions of the modified Fokker-Planck equation as function of the beam energy are compared  with the experimental results. It appears that for the real undulator a bending radius $R$ = 10 mm might be a better approximation. However, the experimental finding in references \cite{BacK12, BacK12A} that only a few percent of all electrons contribute to the coherent part of the peak strongly suggests that only parts of the undulator exhibit the designed structure.

\section{Scaling behaviour of the modified Fokker-Planck equation}\label{scaling}
\begin{figure}[tb]
%\vspace*{-1.1cm}
%\centering
    \begin{minipage}[t]{0.5\linewidth}
    %\centering
    \vspace*{-4.95cm}
     \includegraphics[angle=0,scale=0.57,clip]{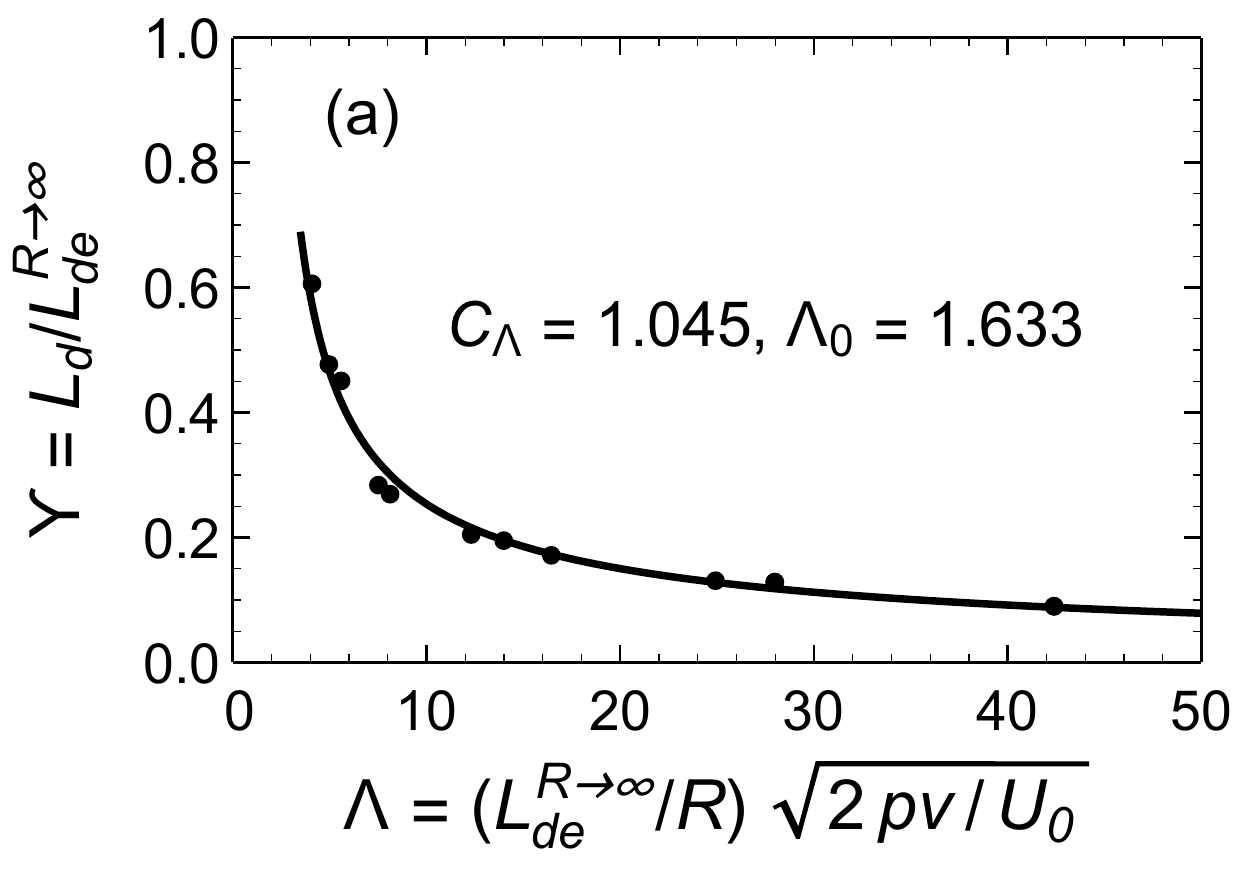}
    \end{minipage}
    \hfill
    %\vspace*{1.1cm}
    \hspace{-2.5cm}
    \begin{minipage}[t]{0.5\linewidth}
    %\vspace*{-5.6cm}
    %\centering
    \includegraphics[angle=0,scale=0.55,clip]{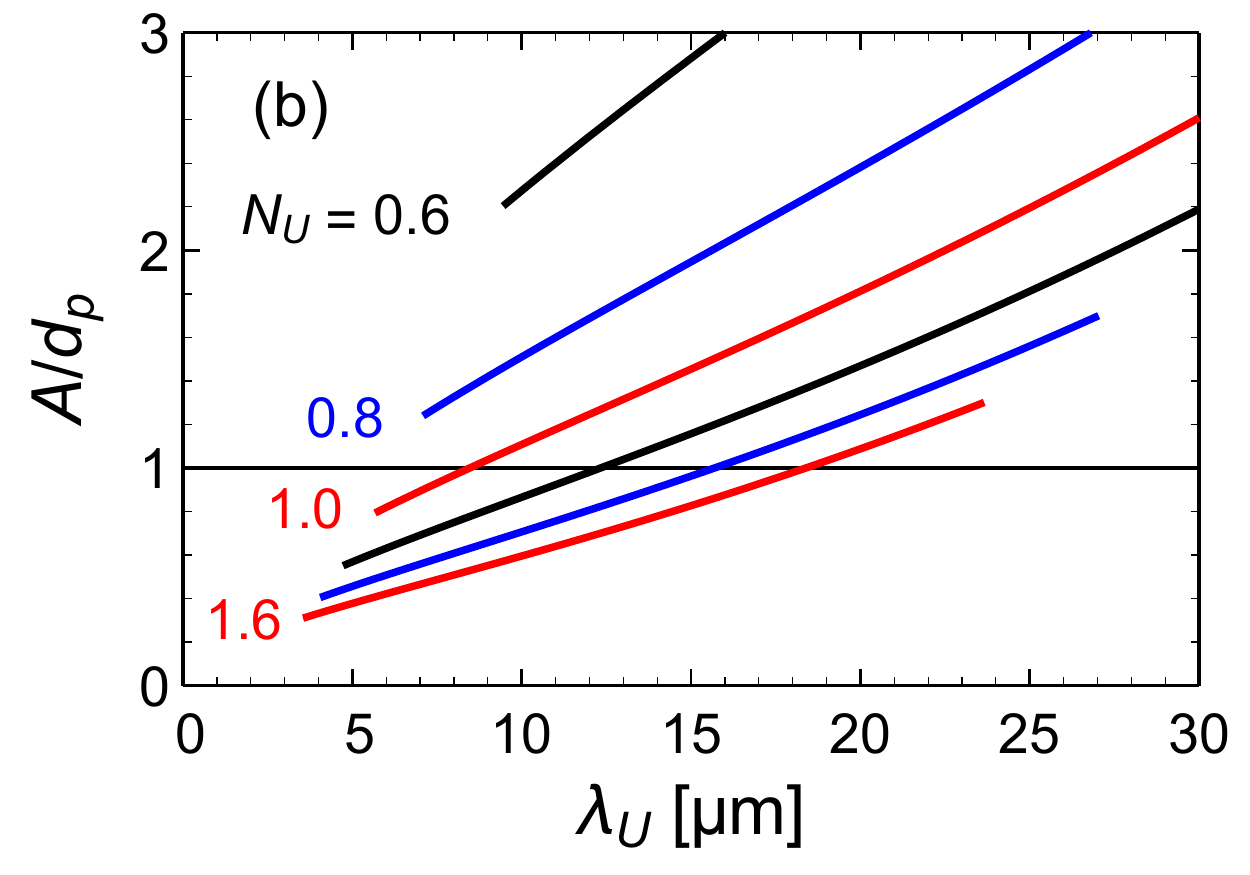}
    \end{minipage}
%\vspace*{-2.1cm}
\caption[]{(a) Dechanneling length $L_{d}$ as obtained from various numerical solutions of the Fokker-Planck equation depicted in normalized variables as defined by equation (\ref{varnorm}). The black curve is a best fit with equation (\ref{FitLdechScalingSi110}). (b) Amplitude $A/d_p$ normalized to the (110) inter planar distance $d_p$ = 1.919 \AA~ for a silicon single crystal as function of the undulator period $\lambda_U$. Beam energy 855 MeV. The bending radius $R$ has been varied in the limits defined by $0 \leq a_n R_{c}/R < 0.4$ with $a_n$ = 0.696. From the equidistance parameter characteristics for $N_U$ the number of meaningful undulator periods can be estimated. For a large amplitude undulator $A/d_p \geq 1$ is required.} \label{LdechScalingSi110}
\end{figure}

After all what has been addressed in the previous sections one might ask the
question what can be expected from an optimized large amplitude undulator
operating with electrons. In principle, this question can be answered solving
the modified Fokker-Planck equation in the parameter space $\lambda_U$,
amplitude $A$, electron energy $E$, and number of periods $N_U$ by a brute-force approach. However, in the following this issue has been approached by
investigating the scaling behaviour of the modified Fokker-Planck equation (\ref{FokkerPlanck1}-\ref{FokkerPlanck3}) which can be rewritten as
\begin{eqnarray}\label{FokkerPlanckScaling01}
\frac{{\partial F(x,{E_ \bot })}}{{\partial x}} + \theta ({E_ \bot } - {U_n})\frac{\partial }{{\partial {E_ \bot }}}\left[\frac{1}{R}\sqrt {2({E_ \bot } - {U_n})pv}\;F(x,{E_ \bot })\right] = \nonumber\\  =  \frac{\partial }{{\partial {E_ \bot }}}\left[ {D_{diff}^{(2)}({E_ \bot })T({E_ \bot })\frac{\partial }{{\partial {E_ \bot }}}\frac{{F(x,{E_ \bot })}}{{T({E_ \bot })}}} \right].
\end{eqnarray}
The same equation with the perpendicular energy variable $E_ \bot$ normalized to $U_0$ and the crystal thickness variable $x$ normalized to the dechanneling length $L^{R\rightarrow\infty}_{de}$ for a bending radius $R\rightarrow\infty$ reads
\begin{eqnarray} \label{FokkerPlanckScaling02}
\frac{{\partial F(x/L_{de}^{R \to \infty },{E_ \bot }/{U_0})}}{{\partial x/L_{de}^{R \to \infty }}} + \theta \left(\frac{E_ \bot -U_n}{U_0}\right)\frac{{L_{de}^{R \to \infty }}}{R}\sqrt {\frac{{2pv}}{{{U_0}}}} \frac{\partial }{{\partial {E_ \bot }/{U_0}}}\left(\sqrt {\frac{E_ \bot -U_n}{U_0}} \;F(x/L_{de}^{R \to \infty },{E_ \bot }/{U_0})\right) = \nonumber \\
= \frac{\partial }{{\partial {E_ \bot }/{U_0}}}\left[ {D_{diff}^{(2)}({E_ \bot }/{U_0})\;T({E_ \bot }/{U_0})\frac{\partial }{{\partial {E_ \bot }/{U_0}}}\left(\frac{{F(x/L_{de}^{R \to \infty },{E_ \bot }/{U_0})}}{{T({E_ \bot }/{U_0})}}\right)} \right] \;\;\;\;\;\;\;
\end{eqnarray}
with
\begin{eqnarray} \label{LdeBaier}
L_{de}^{R \to \infty } = 2\frac{{{U_0}\;pv\;{X_0}}}{{E_s^2}}.
\end{eqnarray}
The latter is exactly the expression which Baier et al. quote \cite[equation (10.1)]{BaiK98}. However, the expression $E_{s,Baier} = \sqrt{2\pi}m/e  = {{15.0\;{\rm{MeV}}}}$ must be replaced for small thicknesses $x \ll X_0$ by our empirical $E_s$ = 10.6 MeV. For Si (110) the barrier hight can well be approximated in the interval $0 \leq a_n R_{c}/R < 0.4$  by $U_n/U_0 = (1- a_n R_{c}/R)^{3.5}$ with $a_n = 0.696$ and the critical radius $R_{c} = a_c \cdot pv$ with $a_c$ = 1.69 mm/GeV. Introducing normalized variables
\begin{eqnarray} \label{varnorm}
\Upsilon = \frac{L_{d}}{L_{d}^{R \to \infty}} = \frac{L_{de}}{2U_0X_0R_{c}/(a_c E_s^2)}~~~~~~~~~ \mbox{and}~~~~~~~~~ \Lambda  = \frac{{L_{de}^{R \to \infty }}}{R}\sqrt {\frac{{2pv}}{{{U_0}}}} = \frac{\sqrt{8U_0}}{E_s^2}\frac{X_0}{a_c^{3/2}}\frac{R_{c}^{3/2}}{R},
\end{eqnarray}
which both exhibit a functional dependence on the critical radius $R_c$, the dechanneling length can be represented as shown in figure \ref{LdechScalingSi110} (a). The black curve is a best fit with the function
\begin{eqnarray} \label{FitLdechScalingSi110}
\Upsilon(\Lambda)=
\frac{{{L_{d}}(\Lambda )}}{{L_{de}^{R \to \infty }}} =
\frac{C_\Lambda}{{{{(\Lambda  - {\Lambda _0})}^{2/3}}}}.
\end{eqnarray}
The parameters $C_\Lambda$ and $\Lambda_0$ are quoted in figure \ref{LdechScalingSi110} (a). Within the limits $4 \lesssim \Lambda \lesssim 50$ of the validity of this approximation the dechanneling length can be calculated with equation (\ref{FitLdechScalingSi110}) for arbitrary combinations of $pv$, and $R$ at fixed $U_0 = 22.56$ eV, $X_0 = 0.0936$ m, and $E_s = 10.6$ MeV.
%%%%%%%%%%%%%%%%%%%%%%%%%%%%%%%%%%%%%%%%%%%%%%%%%%%%%%%%%%%%%%%%
%%%%%%%%%%%%%%%%%%%%%%%%%%%%%%%%%%%%%%%%%%%%%%%%%%%%%%%%%%%%%%%%

\section{Discussion and conclusion}\label{discussion}
With the parametrization described in the previous section \ref{scaling} one
might now be able to seek for optimal parameters for a large amplitude
undulator as follows. The undulator period $\lambda_U$, maximum number of
periods $N_U$, and the dechanneling length $L_{d}(\Lambda)$ should obey the
equation $N_U \lambda_U = 2 L_{d}(\Lambda)$. From simple geometrical
considerations one obtains for the amplitude $A \cong (\lambda_U/4)^2/2R$.
Combining both equations results in the tuple
\begin{eqnarray} \label{AdpUnequ}
\Bigg(\lambda_U =\frac{2 L_{d}\big(\Lambda(pv,R)\big)}{N_U},~~~~ \frac{A}{d_p}=\frac{1}{8}\frac{{L_{d}^2\big(\Lambda(pv,R)\big)}}{{R\;{d_p}N_U^2}} \Bigg)
\end{eqnarray}
which is shown in figure \ref{LdechScalingSi110} (b) for $pv \cong pc$ = 855 MeV, corresponding to $R_c$ = 1.445 mm, with $N_U$ as a parameter. The bending radius $R$ has been varied in the limits defined by $0 \leq a_n R_{c}/R < 0.4$. The requirement $A/d_p \geq 1$ for a large amplitude undulator, $d_p$ = 1.919 \AA~ is the (110) inter planar distance of a silicon single crystal, shows that the coherence length in the investigated parameter space of $\Lambda(pv,R)$ amounts to less than two periods. In other words, neglecting rechanneling it makes no sense to produce at $\lambda_U = 9.9~ \mu$m an undulator with more than about two periods, and at $\lambda_U = 20~ \mu$m with more than about three periods. However, since rechanneling enhances the intensity, many more periods may be advantageous. Although one cannot tell at this stage of the analysis something about the intensity of the emitted undulator radiation, it seems to be quite probable that the design values for the above analyzed $\lambda_U = 9.9~ \mu$m four-period undulator with $A/d_p = 2.42$ were not optimal since the coherence length amounts to only 0.57 periods.

An increase of the beam energy may be beneficial although the dependence on the beam energy turned out to be rather weak. For instance, at 14 GeV the coherence length might be at $A/d_p = 1$ and $\lambda_U = 150~\mu$m close to four periods. The associated undulator parameter is $K = 2 \pi \gamma A/\lambda_U $ = 0.22. Under these circumstances an undulator with 10 periods or more, if advantage will be taken on rechanneling, might emit rather intense radiation with a photon energy of about 12.4 MeV at on-axis observation. Such a device may well be worth for further investigation.
%, perhaps aiming as in reference \cite{UggW15} also at Nuclear Waste %Transmutation.

\acknowledgments

%This is the most common positions for acknowledgments. A macro is
%available to maintain the same layout and spelling of the heading.
Support by the European Commission (the PEARL Project within the H2020-MSCA-RISE-2015 call, GA 690991) is gratefully acknowledged.

%\paragraph{Note added.} This is also a good position for notes added
%after the paper has been written.


%\bibliographystyle{plain}
%\bibliographystyle{elsarticle-num}
%\bibliographystyle{ieeetr}
%\bibliographystyle{unsrt}
\bibliographystyle{JHEP}
\bibliography{bibfileRREPS17}

\end{document}